\documentclass[conference, 10pt]{IEEEtran}
\IEEEoverridecommandlockouts

\usepackage{preamble}
\title{Weakly-Convex Regularization for\\Magnetic Resonance Image Denoising\vspace{-20pt}}
\author{
\IEEEauthorblockN{Akash Prabakar$^{1,2}$, Abhishek Shreekant Bhandiwad$^{1}$, Abijith Jagannath Kamath$^{1}$ and Chandra Sekhar Seelamantula$^{1}$}
\IEEEauthorblockA{$^1$Department of Electrical Engineering, Indian Institute of Science, Bengaluru, India 560 012\\$^2$Voxelgrids Innovations Pvt. Ltd., Bengaluru, India 560 058}
Email: {\selectfont\sffamily \{pakash, abhishekbs, abijithj, css\}@iisc.ac.in}\\
    	\thanks{A portion of this work was accepted at the IEEE International Conference on Acoustics, Speech and Signal Processing (ICASSP) 2025 \cite{kamath2025design}.

    This work is supported by the Anusandhan National Research Foundation (ANRF) under the SUPRA scheme (SPR/2023/00041), Government of India.}
\vspace{-20pt}}
\begin{document}
\maketitle
\begin{abstract}
Regularization for denoising in magnetic resonance imaging (MRI) is typically achieved using convex regularization functions. Recently, deep learning techniques have been shown to provide superior denoising performance. However, this comes at the price of lack of explainability, interpretability and stability, which are all crucial to MRI. In this work, we present a constructive approach for designing weakly-convex regularization functions for MR image denoising. We show that our technique performs on par with state-of-the-art denoisers for diffusion-weighted MR image denoising. Our technique can be applied to design weakly-convex convolutional neural networks with prototype activation functions that impart interpretability and are provably convergent. We also show that our technique exhibits fewer denoising artifacts by demonstrating its effect on brain microstructure modelling.
\end{abstract}
\section{Introduction}\label{sec:intro}
Regularization for image denoising is an important problem in signal/image processing, computational imaging and machine learning. In recent literature, powerful denoisers are deployed in a plug-and-play fashion to solve linear inverse problems \cite{venkatakrishnan2013plug,kamilov2023plug,ryu2019plug,zhang2021plug}. Such techniques may be data-driven while being interpretable, stable and possessing theoretical convergence guarantees, which are crucial in critical applications such as magnetic-resonance imaging (MRI).\\
\indent Recently, convex-non-convex formulations have been shown to provide performance superior to that of convex formulations for regularized inverse problems \cite{lanza2021convex,goujon2024learning,shumaylov2024weakly,nareddy2022ensemble}. Amongst nonconvex regularization functions, weakly-convex functions are preferred as it is possible to provide convergence guarantees \cite{pokala2019firmnet,pokala2021iteratively}.\\
\indent In this work, we consider regularization for denoising and solve problems of the type:
\begin{equation}\label{p:canonical-denoiser}
    \minimize{\bd x\in\bb R^n}\;\frac12\norm{\bd y-\bd x}_2^2 +  g(\bd x),
\end{equation}
where $g:\bb R^n\rightarrow\bb R$ denotes a weakly-convex regularizer.
\begin{definition}\label{def:weakly-convex}
Let $\rho>0$. A function $g:\bb R^n\rightarrow\bb R$ is $\rho$-weakly convex if $g+\frac{\rho}{2}\norm{\cdot}^2$ is convex.
\end{definition}
Observe that the objective function in \eqref{p:canonical-denoiser} is convex whenever $g$ is $\rho$-weakly-convex with $\rho<1$ \cite{bauschke2011convex,parikh2014proximal,beck2017first,goujon2024learning}.
\begin{figure}[!t]
\resizebox{3.4in}{!}{
\begin{tikzpicture}
    \node[anchor=east](A) at (0,0)
    {\includegraphics[width=0.34\linewidth]{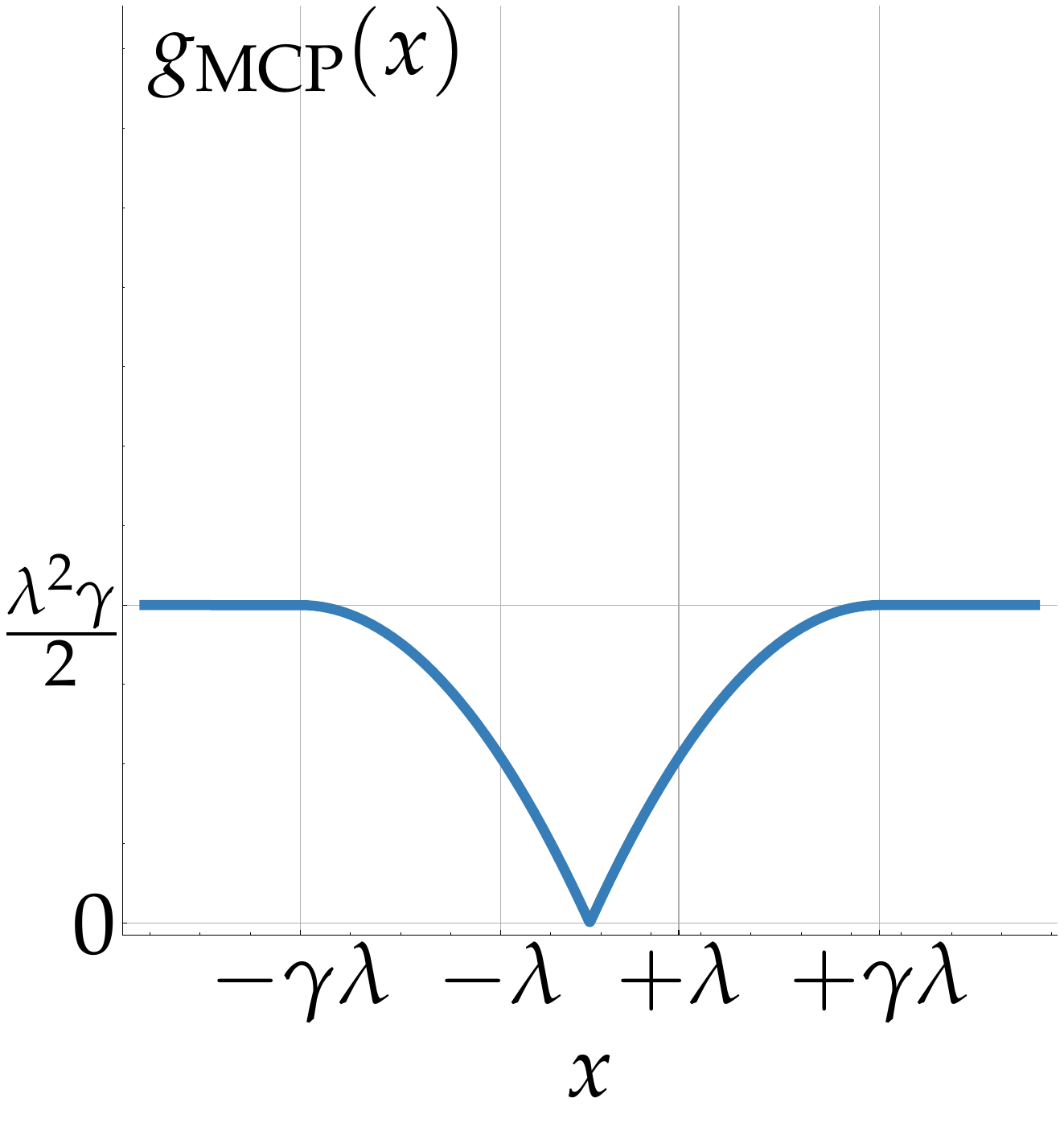}};

    \node[anchor=east](B) at (2.8,0)
    {\includegraphics[width=0.3\linewidth]{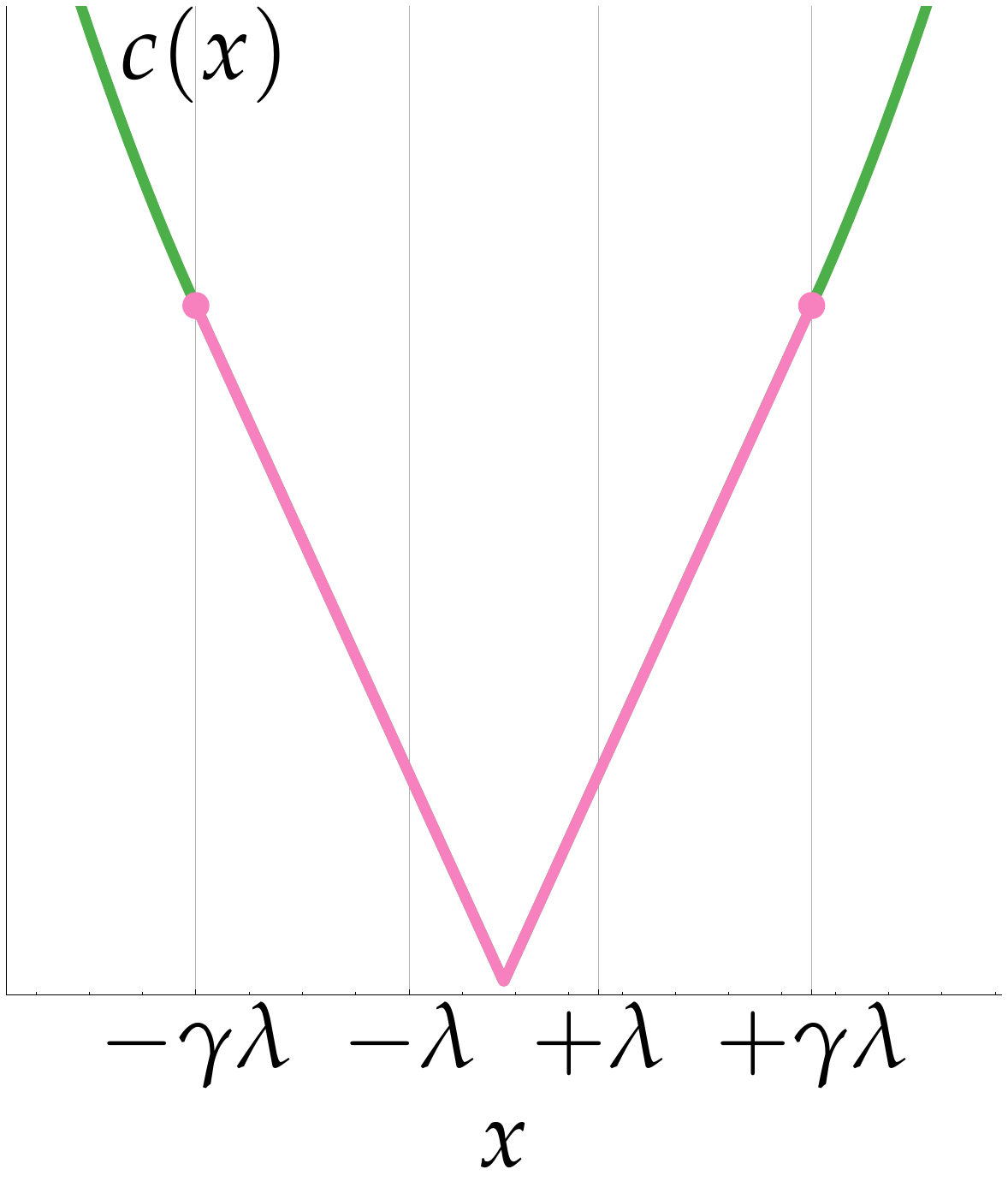}};
    \node[anchor=east](A1) at (.05,0.225){$=$};

    \node[anchor=east](C) at (5.7,0)
    {\includegraphics[width=0.3\linewidth]{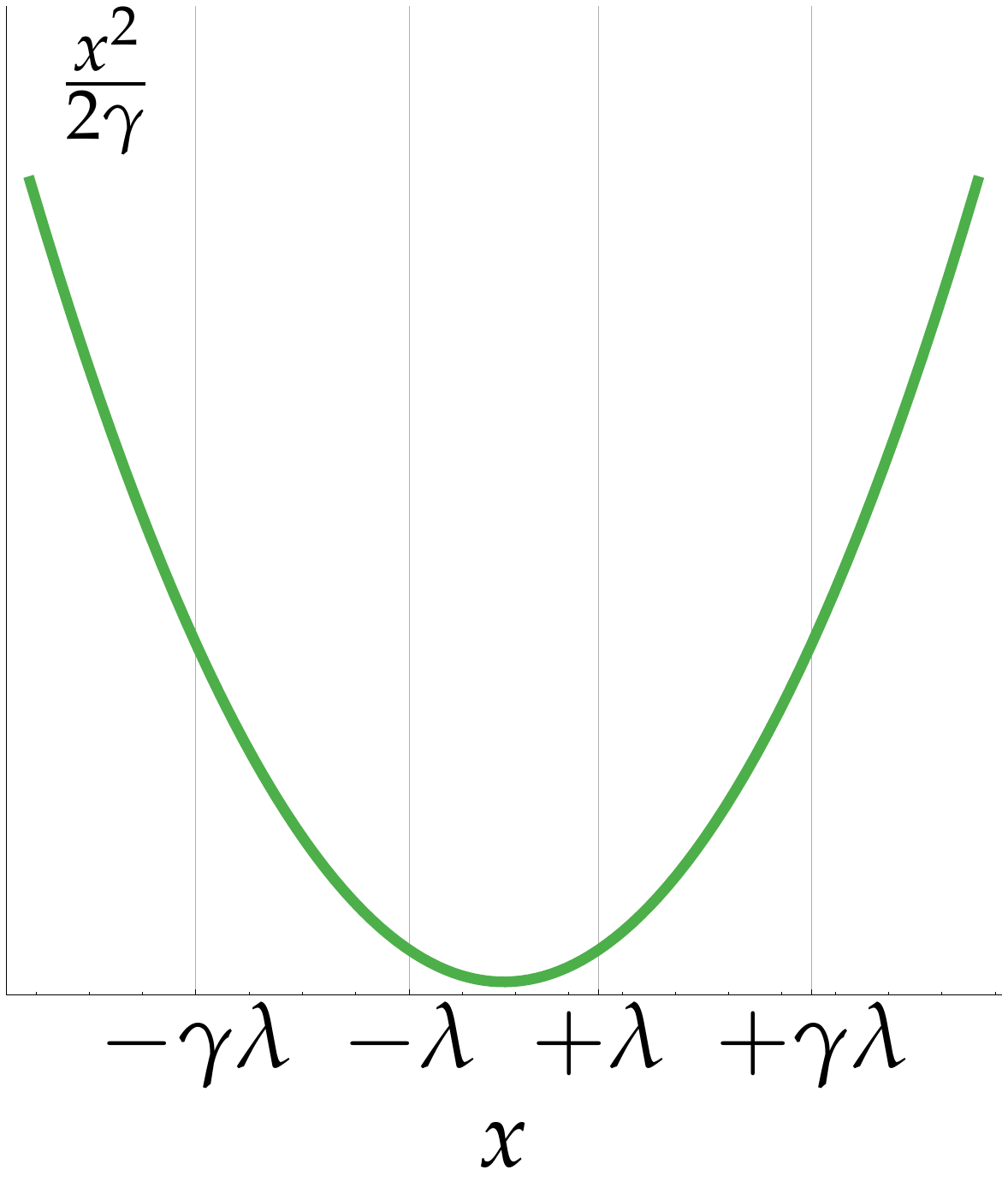}};
    \node[anchor=east](B1) at (2.85,0.225){$-$};
\end{tikzpicture}}
\caption{Constructing the minimax-concave penalty (MCP) as the difference between the convex function $c(x)$ and the quadratic function $\frac{x^2}{2\gamma}$.}
\label{fig:mcp-construction}
\end{figure}
\begin{figure}[!t]
    \centering
    \includegraphics[width=.95\linewidth]{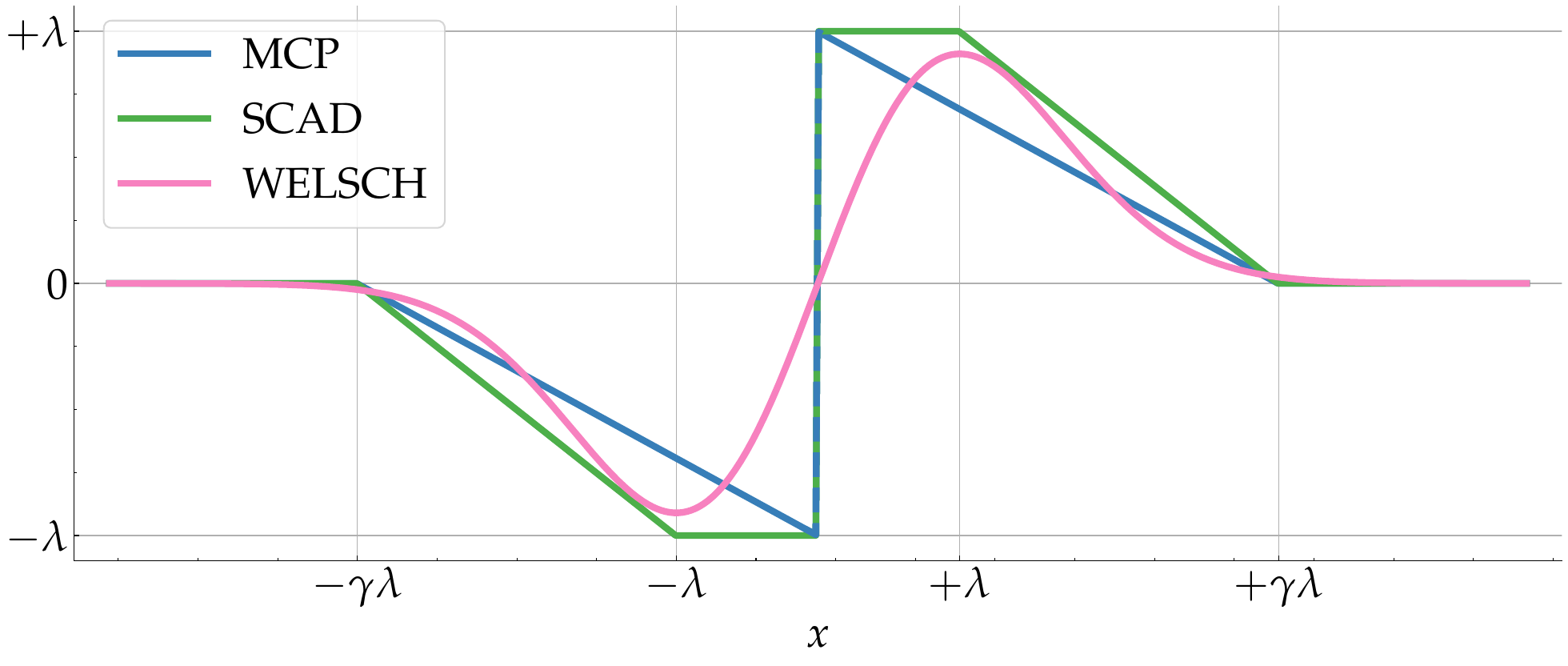}
    \caption{Weakly-monotone derivatives of weakly-convex functions MCP, SCAD and the Welsch penalty.}
    \label{fig:activations}
\end{figure}
\begin{figure*}[!t]
    \centering
    \includegraphics[width=\linewidth]{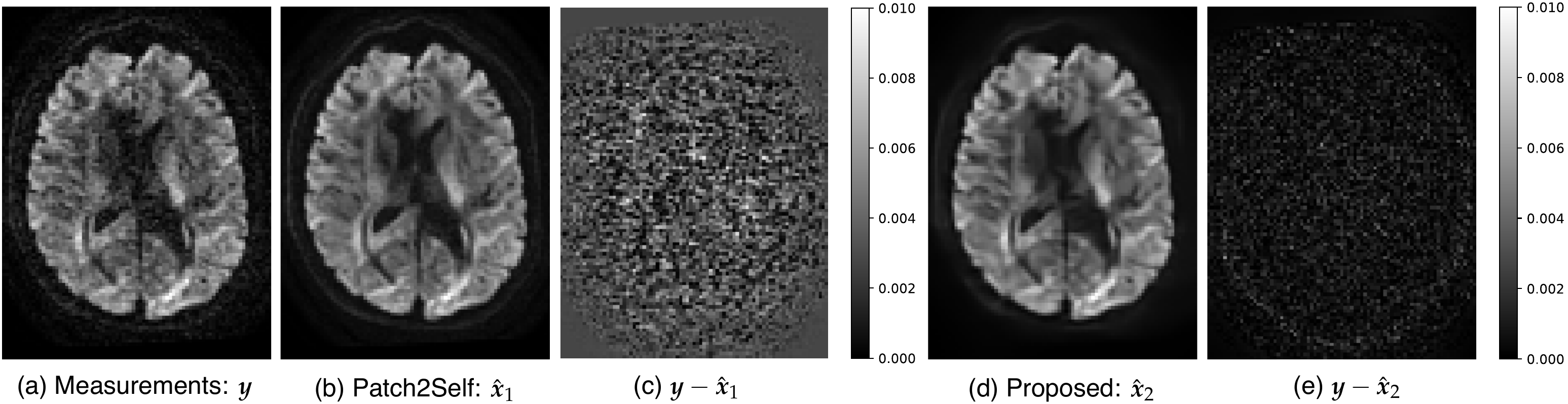}
    \caption{Denoising diffusion weighted MR image $\bd y$ using Patch2Self $\hat{\bd x}_1$ and weakly-convex ridge regularizer network with the Welsch function as activation.}
    \label{fig:dwi_denoising}
\end{figure*}
\section{Construction of Weakly-Convex Functions}\label{sec:construction}
Several works consider the design of weakly-convex functions for regularized inverse problems. Recently, Goujon {\it et al.} considered the family of piecewise-linear functions that are learnt within a ridge-regularized network \cite{goujon2024learning}. Shumaylov {\it et al.} considered input-weakly-convex networks within the unsupervised setting \cite{shumaylov2024weakly}. We proposed a bottom-up approach for construction of regularizers using prototype scalar-input weakly-convex functions to build weakly-convex convolutional networks \cite{kamath2025design}. We present two methods to design such prototype functions.
\subsection{Minimizing A Measure of Concavity}
For a nonconvex function $\psi:[0,+\infty)\rightarrow [0,+\infty)$, we can define the measure of maximum concavity as
\[
    \kappa(\psi) = \sup_{0<x_1<x_2} \frac{{\psi'}(x_2)-{\psi'}(x_1)}{x_1-x_2}.
\]
Nontrivial minimization of $\kappa(\psi)$ requires boundary conditions. The simplest example that is continuously differentiable in $(0,+\infty)$ arises from using a single knot point at $x=\gamma\lambda$:
\begin{align*}
    \psi(x) = \begin{cases}
        \lambda x-\dfrac{x^2}{2\gamma}, & 0\leq x\leq\gamma\lambda, \\
        \dfrac{\lambda^2\gamma}{2}, & \gamma\lambda<x,
    \end{cases}
\end{align*}
with $\kappa(\psi)=1/\gamma$. The minimax-concave penalty (MCP) $g_{\text{MCP}}:\bb R\rightarrow\bb R$ is obtained by symmetrizing $\psi$ as $g_{\text{MCP}}(x)=\psi(\abs{x})$ \cite{zhang2010nearly,pokala2021iteratively}. The MCP can be written as a difference of a convex function and a quadratic function:
\begin{align}
	\intertext{}
	g_{\text{MCP}}(x) = \eqnmarkbox[blue]{cvx}{c(x)} - \eqnmarkbox[red]{quad}{\frac{x^2}{2\gamma}}.
    \annotate[yshift=.5em]{above,left}{cvx}{Convex}
    \annotate[yshift=.5em,xshift=.75em]{above,right}{quad}{Quadratic}
\end{align}
as illustrated in Fig.~\ref{fig:mcp-construction}. In this way, it is easy to see that the MCP is weakly convex with $\rho=1/\gamma<1$, $\gamma>1$. Similarly, setting two knot points at $x=\lambda$ and $x=\gamma\lambda$, we get the smoothly-clipped absolute deviation (SCAD) penalty \cite{fan2001variable}. In general, we have the following result from \cite{kamath2025design}.
\begin{theorem}\label{thm:main-result}
	Suppose $\psi:[0,+\infty)\rightarrow[0,+\infty)$, and let $g:\bb R\rightarrow [0,+\infty)$ be defined as $g(x)=\psi(\abs{x})$. Then, $\kappa(\psi)=\rho$ if and only if $g$ is $\rho$-weakly convex.
\end{theorem}
\subsection{Smooth Functions with Bounded Second Derivative}
From Definition~\ref{def:weakly-convex}, it is verifiable that a smooth function is weakly convex if its second derivative is bounded below.
\begin{theorem}[cf.~\cite{goujon2024learning}]
    Let $g:\bb R\rightarrow\bb R$ be differentiable with $L$-Lipschitz derivative. Then, $g$ is $\rho$-weakly convex with $\rho\leq L$.
\end{theorem}
The class of smooth functions includes commonly used activation functions in modern deep learning frameworks and activations that have been shown to be effective in representation learning, such as Welsch activation \cite{dennis1978techniques}, sinusoidal activation \cite{sitzmann2019siren}, Gabor wavelet activation \cite{saragadam2023wire}. Fig.~\ref{fig:activations} shows the weakly-monotone derivatives of the MCP, SCAD and the Welsch penalty that are used in a ridge regularizer network to construct weakly-convex regularization functions for denoising.
\section{Denoising Diffusion MR Images}\label{sec:experiments}
\begin{figure}[!t]
    \centering
    \includegraphics[width=\linewidth]{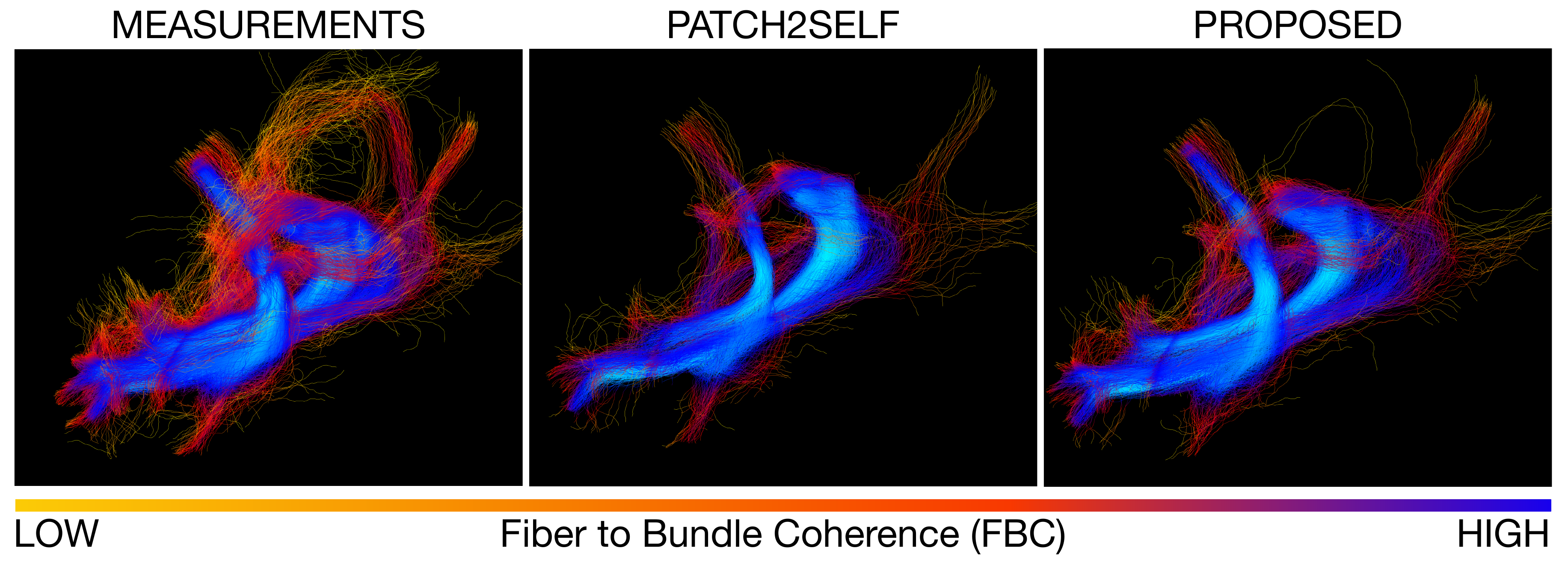}
    \caption{Fiber to bundle coherence (FBC) density maps \cite{portegies2015improving} obtained using a probabilistic tracking algorithm. The proposed denoising technique retains the coherent streamlines while removing incoherent streamlines.}
    \label{fig:FBC}
\end{figure}
We demonstrate our denoising technique on diffusion-weighted MR images \cite{basser1994mr} from the Stanford HARDI dataset \cite{rokem2013high}. We pose the denoising problem as in \eqref{p:canonical-denoiser} with $g(\bd x)$ as a convolutional ridge regularizer with Welsch activation functions. We unfold the iterations of the subgradient descent algorithm to obtain a deep-unfolded network \cite{monga2021algorithm}. Details of training the network are given in \cite{kamath2025design}.\\
\indent Fig.~\ref{fig:dwi_denoising} shows measurements $\bd y$, the image denoised using the Patch2Self technique \cite{fadnavis2020patch}, denoted $\hat{\bd x}_1$, and the image denoised using the proposed technique, denoted $\hat{\bd x}_2$. We note that, in certain areas, Patch2Self removes structure from the measurements, whereas our technique shows fewer artifacts. This observation is more prominent in the bright pixels of the residuals denoted $\bd y-\hat{\bd x}_k,\;k=1,2$.\\
\indent Further, we evaluate our denoising technique by performing tractography on diffusion tensor measurements obtained from the Stanford HARDI dataset. Fig.~\ref{fig:FBC} shows fiber to bundle coherence (FBC) density maps\cite{portegies2015improving}. The proposed technique retains coherent streamlines more accurately compared to Patch2Self, while removing incoherent streamlines.
\section{Conclusions}\label{sec:conclusions}
We considered a bottom-up approach for constructing weakly-convex regularization functions for MR image denoising. The proposed technique is provably convergent, interpretable, data-driven and demonstrates denoising with fewer artifacts compared to state-of-the-art deep-learning-based techniques for MR image denoising. We validate the claim by performing tractography on the diffusion tensor measurements.
\balance
\bibliographystyle{ieeetr}
\bibliography{refs.bib}

\end{document}